# Haptic Dial based on Magnetorheological Fluid Having Bumpy Structure

Seok Hun Lee, Yong Hae Heo, Seok-Han Lee, and Sang-Youn Kim
*Interaction Laboratory, Korea University of Technology and Education, Cheonan, South Korea*
{seokhoon99, huice, shlee, sykim}@koreatech.ac.kr

**Abstract** --- We proposed a haptic dial based on magnetorheological fluid (MRF) which enhances performance by increasing the MRF-exposed area through concave shaft and housing structure. We developed a breakout-style game to show that the proposed haptic dial allows users to efficiently interact with virtual objects.

Keywords: haptic dial, magnetorheological fluid, kinesthetic actuator, torque feedback

## 1 INTRODUCTION

Dial interfaces offer more precise control compared to other input interfaces, such as buttons or switches, which are typically used for simple "on/off" functions. One of the most critical factors in designing input devices is providing users with timely and useful feedback through haptic sensations. Recently, magnetorheological (MR) fluids have been used to create compact and miniaturized haptic modules [1-3]. To maximize the resistive torque generated by MR fluids in haptic modules, new structures that simultaneously engage multiple operating modes have been proposed [4-6]. In this paper, we present a haptic dial that provides sufficiently large resistive torque to the user.

## 2 DESIGN OF THE PROPOSED HAPTIC DIAL

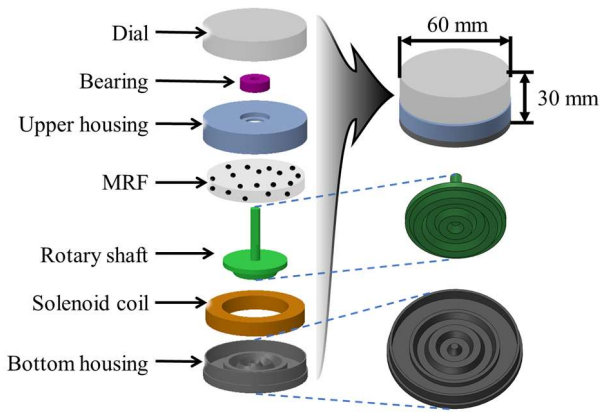

Fig.1  Structure of the proposed haptic dial

The proposed haptic dial consists of a rotator, a bearing, an upper housing, a magnetorheological fluid (MRF) [7], a rotary shaft, a solenoid coil and a lower housing, as shown in Figure 1. The more the amount of MRF activated by the magnetic field, the better the haptic dial performs. Since we need to maximize the magnetically active area of the magnetorheological fluid between the rotary shaft and the bottom housing, the rotary shaft and bottom housing are designed with a bumpy structure to increase the area in contact with the magnetorheological fluid.

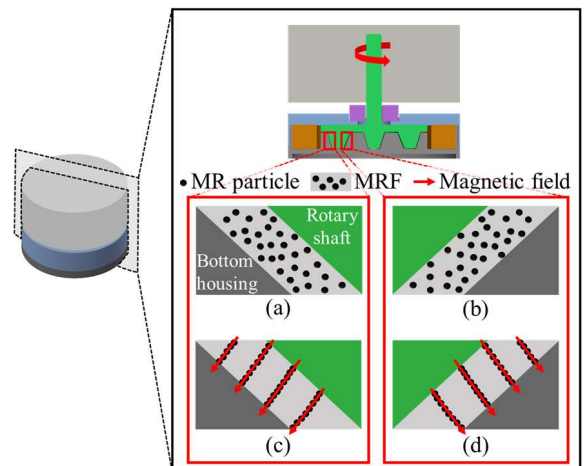

Fig.2  The operating principle of proposed haptic dial. The MRF between rotary shaft and bottom housing at initial state ((a) tilted left region and (b) tilted right region) and magnetic field on state ((c) tilted left region and (d) tilted right region).

Figure 2 shows operating principle of the proposed haptic dial. The MRF acts as a lubricant which allows the shaft to rotate smoothly in initial state, as shown in Figures. 2(a) and 2(b). Under magnetic field is applied to the MRF, iron particles are aligned according to the magnetic field as shown in Figures 2(c) and 2(d). The aligned iron particles increase the resistive force when the user rotates the haptic dial.

## 3 VIRTUAL ENVIRONMENT

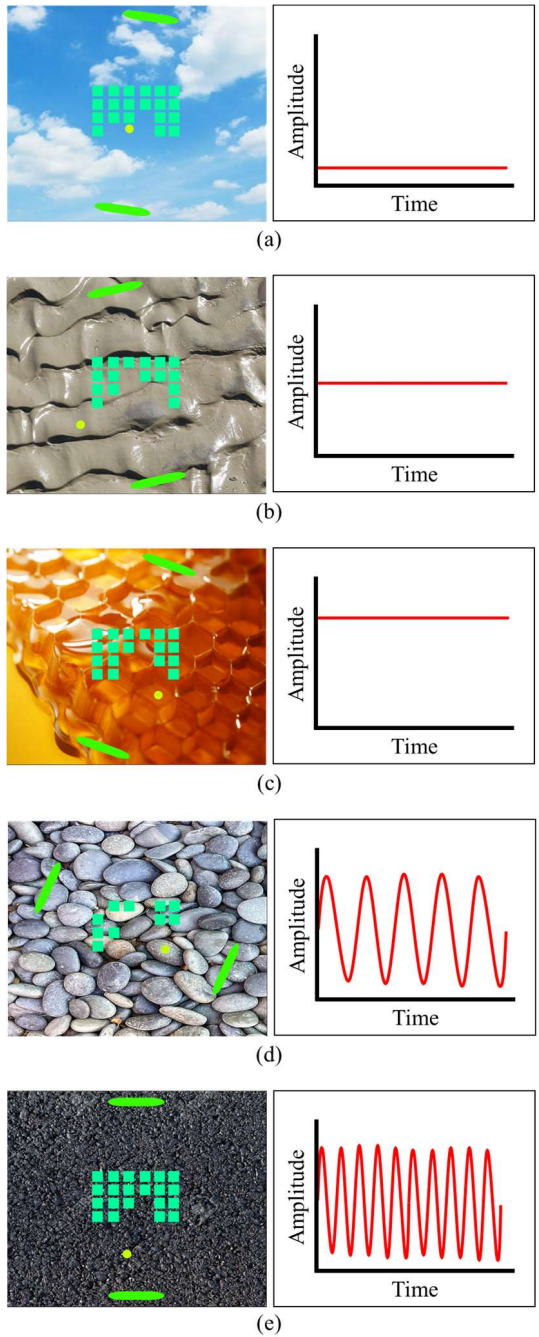

Fig.3 The virtual environment of the brick out game with different virtual background and the input control signal for the haptic dial ((a): sky, (b): mud, (c): honey, (d): pebble, (e): asphalt)

We implemented a breakout-style game to verify the performance of the proposed haptic dial as shown in Figure 3. The user can enjoy the game while feeling different haptic sensations based on the game background. When the background is sky, the user feels weak resistive torque; when it is mud, the user feels strong resistive torque; and when it is honey, the user feels very strong resistive torque. Additionally, the user feels rough vibrations when the background is filled with pebbles, and very fine vibrations when the background is asphalt.

## 4 CONCLUSION

In this paper, we proposed haptic dial based on magnetorheological fluid (MRF) with a bumpy structure. To enhance the performance of the proposed haptic dial, we maximized the area where the MR fluid is activated by designing the shaft and housing with a bumpy structure. We also developed breakout-style game with varying virtual backgrounds enabling users to experience diverse haptic sensations. We expect that the proposed haptic dial shows potential for use in applications such as human-computer interfaces, gaming, and education.

## ACKNOWLEDGEMENT

This work was supported by Priority Research Centers Program through the National Research Foundation of Korea (NRF) funded by the Ministry of Education (NRF-2018R1A6A1A03025526). This work was also supported by the Technology Innovation Program (RS-2024-00419333, Development of lead-free piezoelectric materials and actuator module technology that can be implemented simultaneously with hearing and touch for OLED panels) funded By the Ministry of Trade, Industry & Energy(MOTIE, Korea). We thank for the Cooperative Equipment Center at Koreatech for assistance.